\documentclass[11pt,twoside]{article}
\usepackage{asp2004}
\usepackage{psfig}
\usepackage{epsf}
\usepackage{graphics}
\usepackage{lscape}
\def\edcomment#1{\iffalse\marginpar{\raggedright\sl#1\/}\else\relax\fi}
\reversemarginpar

\begin{document}
\title{Pulsar Physics at Low Frequencies}
\author{ J. A. Eilek$^1$, T. H. Hankins$^1$ and  A. Jessner$^2$}
\affil{$^1$ New  Mexico Tech, Socorro NM 87801 USA}
\affil{$^2$ Max-Planck Institut f\"ur RadioAstronomie, Bonn, Germany}

\begin{abstract}

Recent work has made it clear that the ``standard model'' of pulsar
radio emission cannot be the full answer.   Some fundamental
assumptions about the magnetic field and plasma flow in the radio-loud
region have been called into question by recent observational and
theoretical work, but the solutions to the problems posed are far from
clear. 
It is time to formulate and carry out new observational campaigns 
designed to address these problems; sensitive  low-frequency 
observations will an important part of such a campaign. Because pulsars 
are strong at low frequencies, we believe there will be a good number of 
candidates even for high-time-resolution single pulse work, as well as 
mean profile and integrated spectrum measurements. Such data can push 
the envelope of current models, test competing theories of the radio 
loud region, and possibly provide direct measures of the state of the
 emitting plasma.

\end{abstract}
\thispagestyle{plain}

\section{Introduction}

Pulsar radio emission is not well understood.  Much current work is
grounded in standard models and assumptions about the geometry of the
emitting region and about the nature and dynamics of the plasma
therein.  While these models work well for some stars, we now know they
fail for others.   It has become clear from recent work that specific
predictions of this model are not always met in real stars, and that
important assumptions within the model may  be inconsistent with
the data.    Thus, while the standard model of the pulsar
magnetosphere has served us well, it now seems time to revisit it
critically, with a careful eye to what the data tell us about
the physics in the radio emission region. 

We think that the most fruitful research is the study of  just
those stars in which the standard model fails seriously.  These
will be the stars that reveal the most about conditions in the  
emission region, and help us to formulate the next generation of
models.  In addition, the high quality of modern pulsar data acquisition
systems justifies renewed observational effort: what observations can 
be carried out which will guide and critically test the
models?

Because pulsar emission shows strong frequency evolution, low
frequency observations are an important part of the answer.  However,
although pulsars are strong at low frequencies, good data
below $100$ MHz are hard to come by.  
The LWA thus comes at an opportune time.  In this paper, we discuss
some areas in which the standard model
fails, and then describe three types of low-frequency experiments
which will be important in developing and testing the next generation of pulsar
models. 

\section{The Pulsar Setting}

A ``standard model'' of a pulsar's magnetosphere has been developed
from early  observations and theoretical work.  A rapidly rotating
neutron star   
supports a strong, misaligned, dipolar magnetic field.  The
plasma-filled magnetosphere corotates with the star, except in the
open field line region where the star's magnetic fields connect to the
universe beyond the light cylinder.  Plasma in this open field line
region (somehow) emits coherent radio emission;  we observe ``pulses''
when the star's rotation sweeps this forward-beamed emission region
past our line of sight.      

While this picture is very likely true in general, the devil can be in
the details.   For example, two interesting results which relate
particularly to low frequency observations have emerged from our
recent work.  We now know the geometry of the emission region is more
complex in many stars than had been thought.  We also  strongly
suspect that the plasma in the emission region is much less
dense than had been assumed, and thus must be highly turbulent.

\subsection{Geometry in the Radio Emission Region} 

A fundamental assumption of the standard model is that 
a strong, dipolar magnetic field dominates the plasma
dynamics in the radio emission region.  This picture makes specific,
quantitative predictions.  One 
prediction involves the rotation of the linear polarization angle as
the emission beam sweeps past the observer's sight line.   Another
involves the mean emission profile:  we should
only detect radio emission when the line of sight intersects open
field lines.

In addition, the standard model now includes 
frequency evolution of the mean profile, based on trends observed in
many stars. For instance,  ``radius to frequency mapping'' says
lower-frequency profiles should be broader;  this may connect to
density stratification in the emission region.  Or, profiles should evolve
from ``cone'' emission at high frequencies to ``core'' emission at
low frequencies;  this may reflect different plasma states or emission
mechanisms in different parts of the open field line region.

These predictions are not
always borne out by the data.  We gathered time-aligned mean
profiles and high-quality polarimetry, and used these data to study
the geometry of the emission region in 52 bright pulsars (Hankins \&
Rankin 2005, Eilek \& Hankins 2005).  We did, indeed, verify that the
standard picture works in  about half of the stars --- but in the
other half it does not.  Many stars show complex frequency-dependent
changes in their mean emission profile, which do not follow the
trends predicted by the standard model. Some stars even show emission
at rotation phases which cannot arise from open field lines in a
simple dipole geometry.   In addition, many
stars show striking deviations from the position angle behavior
predicted for a pure dipole magnetic field in the emission region.
These deviations can be steady in some 
stars, fluctuating in others, and can change with observing frequency.
Other stars show evidence that the magnetic 
field switches between two different, quasi-stable states, which is
impossible if the magnetic field is a static dipole.

\subsection{Plasma Density in the Radio Emission Region}

Another fundamental assumption in the standard model is that the
charge density of the radio-loud plasma is exactly that needed to
shield the rotation-induced electric field, and thus to allow the plasma
magnetosphere to corotate smoothly with the star (after Goldreich
\& Julian 1969).  It follows that the plasma in the radio
emission region would be undergoing smooth, coasting flow.  This
assumption is attractive in that it simplifies  
the modelling;  but it seems to be inconsistent with the fact that we
observe pulsars at low radio frequencies.

The connection between plasma density and observing frequency comes
from  the coherent emission mechanism.   Nearly all coherent plasma
emission processes operate at the local (comoving) plasma frequency,
which depends on the local number density of the plasma ({\it e.g.},
Kunzl {\it et al.\,}1998). Our recent
observations of the pulsar in the Crab Nebula support this
hypothesis.  The radio emission in this pulsar is  narrow-band $(0.1 <
\Delta \nu / \nu < 1$) and comes in 
bursts of very short duration (unresolved at 2 nanoseconds;  Hankins
{\it et al.\,}2003).   These observations are consistent with both the
bandwidth and the timescales predicted by Weatherall's
numerical simulations (1997, 1998) of radio emission from collapsing
solitions in strong plasma turbulence,   but they are inconsistent with
the timescales predicted by competing models of the emission
mechanism.  It therefore seems likely that strong plasma turbulence
creates coherent radio emission in the Crab pulsar.  

If plasma emission is indeed the emission mechanism in pulsars, we can
directly determine the  density (in the star's frame) of the emitting plasma
(with some dependence on the  streaming speed of the plasma;  we
follow modern models which find $\gamma \sim 10^2$-$10^3$, {\it
  e.g.\,}Arendt \& Eilek  
2002).  This exercise tells us that the density of the plasma emitting
at, say, 1 GHz or 100 MHz is far below that needed for smooth corotation.  
This fact has two important consequences.

(1) High and low frequency emission must originate
in different parts (high and low density) of the radio-loud region.
In some stars these regions may be stratified in altitude (consistent
with radius-to-frequency mapping in their mean profiles), but in other
stars the high and low density regions appear to  coexist --- as would
be the case in a highly turbulent plasma.  

(2)  The
rotation-induced electric field cannot be fully shielded.  The plasma 
must be subject to strong local forces, and is probably in a 
 highly unsteady, turbulent state.  The lower the density, the more
 unsteady the plasma should be.  
  Because dynamical timescales in this system are on the order
of microseconds, we suspect this unsteady plasma flow is the cause of 
microstructure, which exists in many stars and is especially strong at
low frequencies.

\section{Low Frequency Experiments We Would Like to See}

Broad frequency coverage is crucial to testing theoretical 
models, because the characteristics of pulsar radio emission can
change dramatically between 100 MHz and 10 GHz, and because different
frequencies are emitted by different parts of the radio-loud plasma.
In fact, if our hypothesis of radio emission from strong
plasma turbulence is correct, the strongest discrepancies between data and
models should arise at low frequencies.   
We describe three types of pulsar observations which address these
problems and which we hope could be carried out with the LWA.

\subsection{Integrated Pulsar Spectra}

Time-integrated pulsar fluxes  are the easiest experiment to do, 
because the stars are bright at low frequencies (as in Figure 1).
The spectrum of some stars turns over well above 100 MHz, while that
of others continues rising to the lowest observable frequencies.

\begin{figure}[ht]

\begin{center}
\scalebox{0.212}{\includegraphics{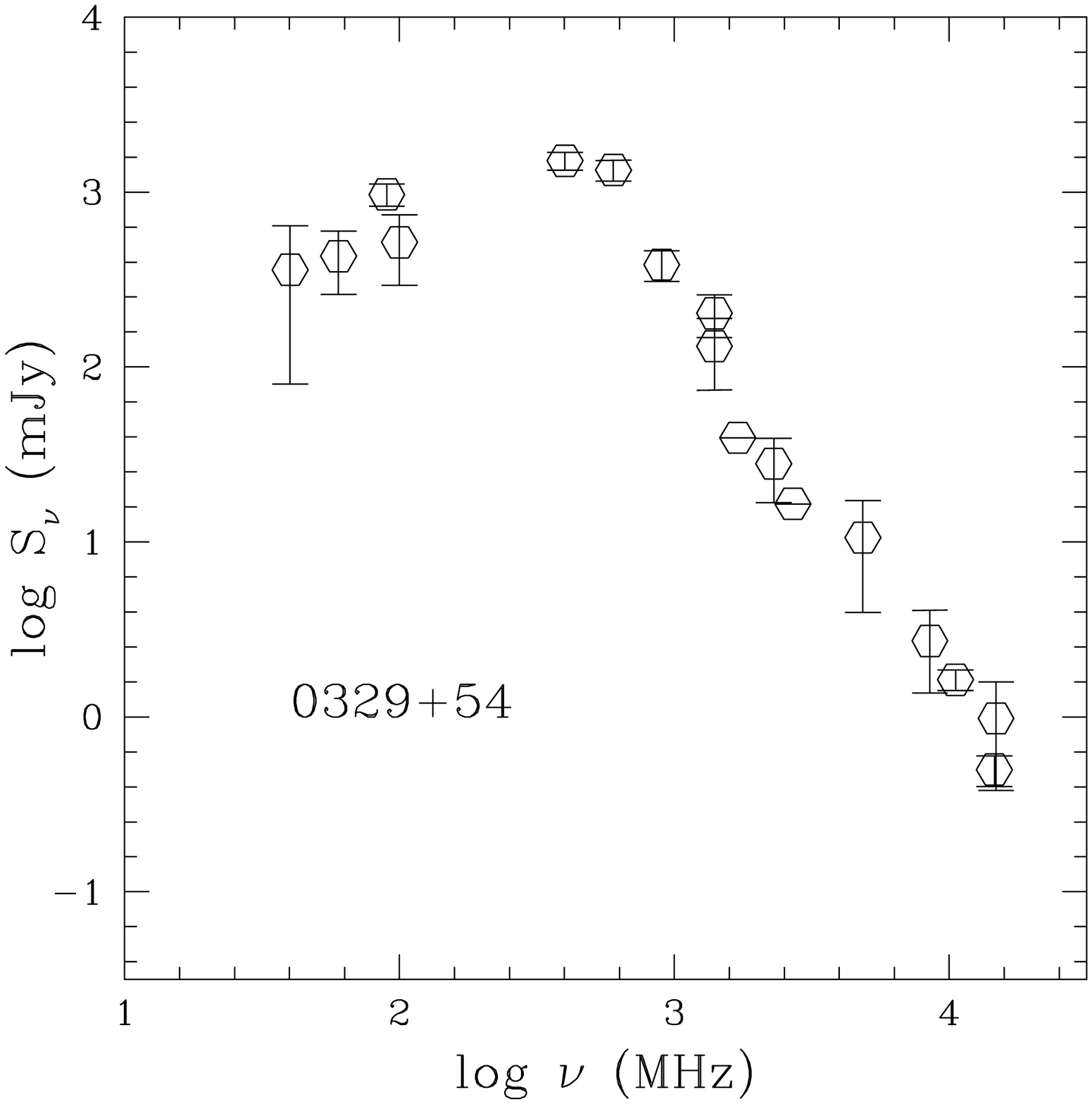}}
\scalebox{0.212}{\includegraphics{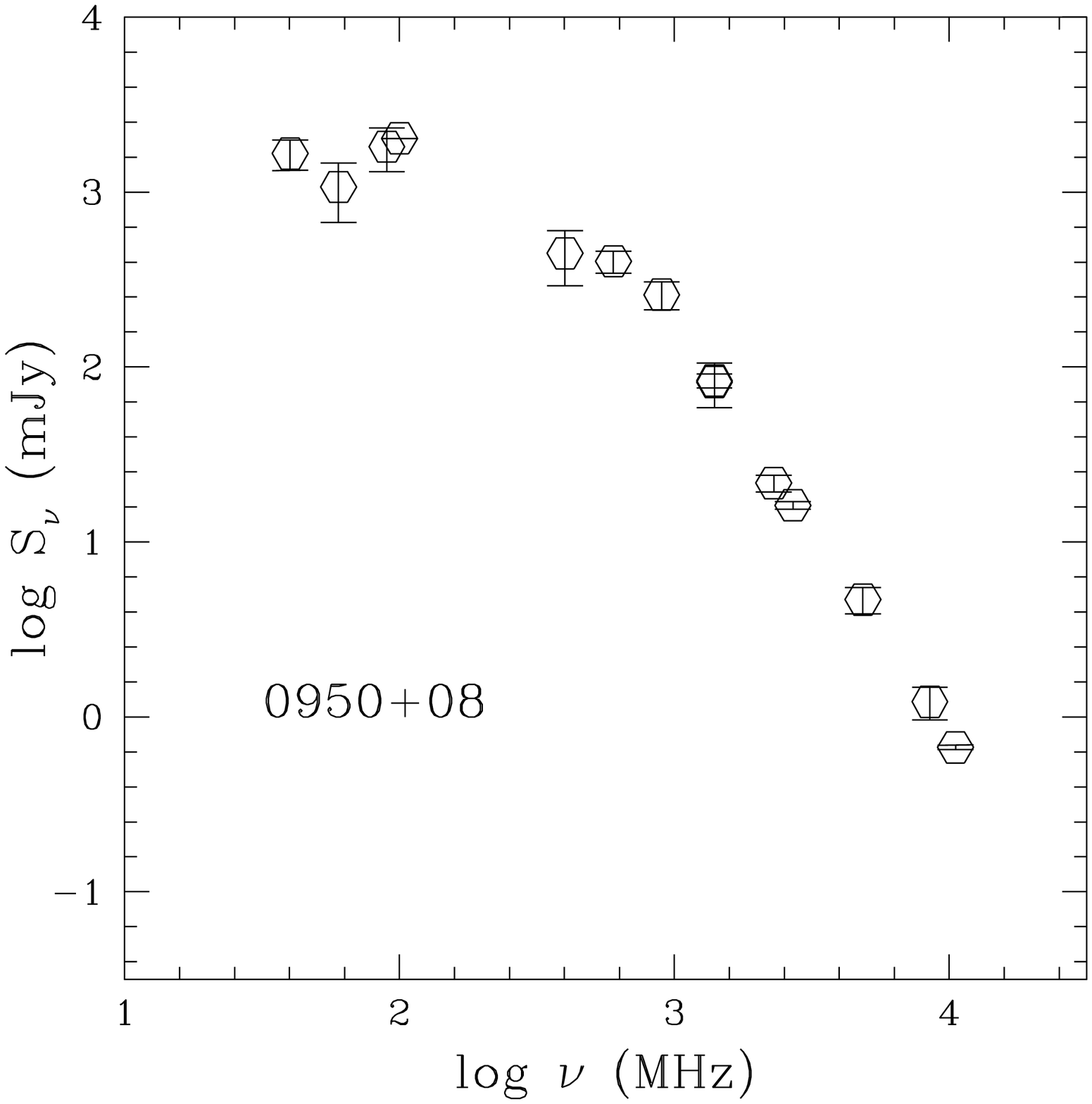}}
\scalebox{0.212}{\includegraphics{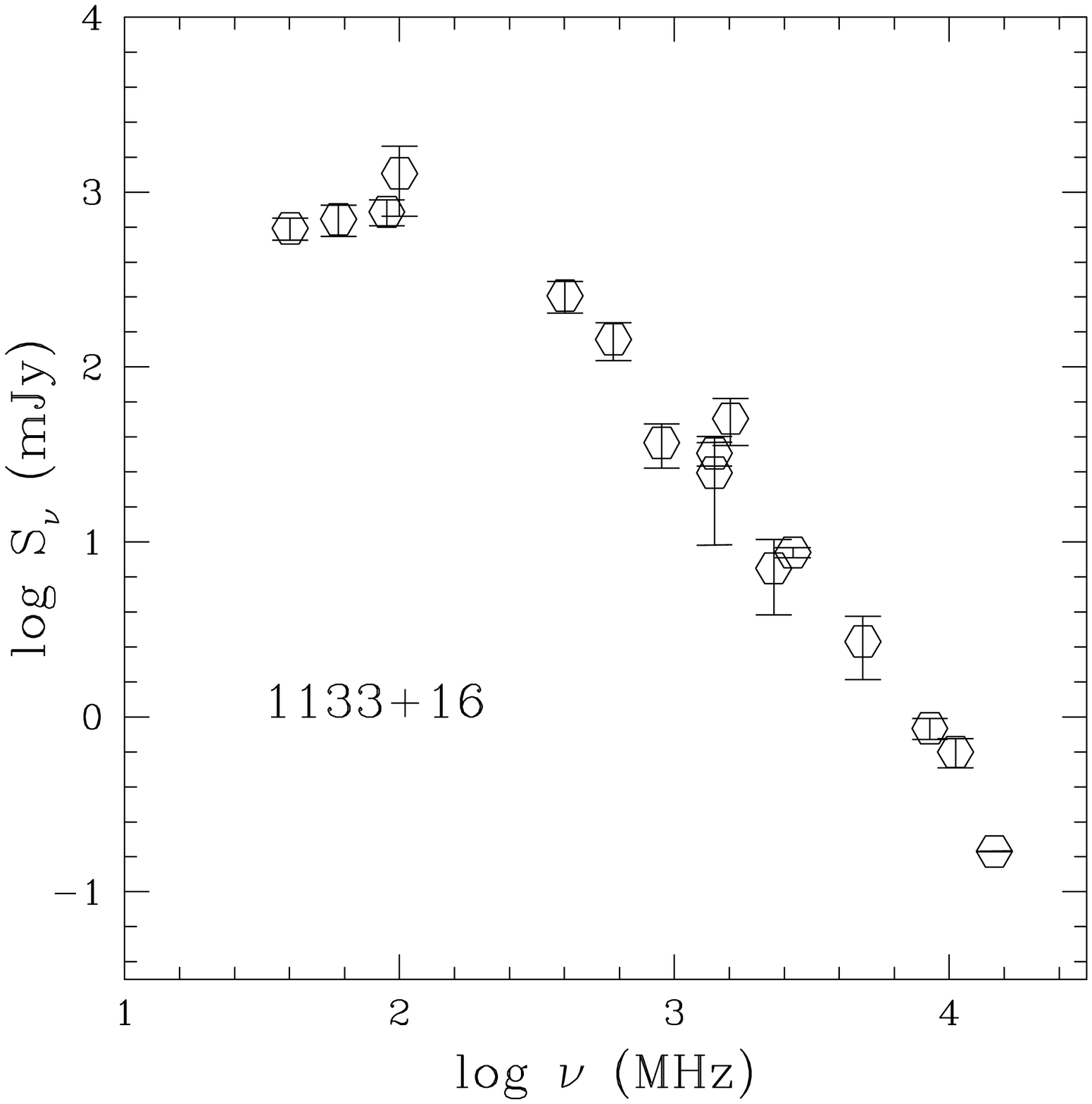}}
\end{center}
\vspace{-2ex}
\caption{Integrated spectra of three bright pulsars, from Malofeev
  {\it et al.\,}   (1994).  The spectrum of some pulsars continues to
  rise to the   lowest observed frequencies, while others turn over at
  several   hundred MHz.  There is as yet no good understanding of
  these spectra.} 
\end{figure}

The integrated spectra are among the hardest data
to interpret, because of the lack of robust spectral predictions by
different emission models. 
Nonetheless, we suggest that probing the spectrum and frequency range of
such emission can be important as a constrant on, or test of, future
models.  This will require LWA-derived fluxes to be 
be combined with higher-frequency data from other 
telescopes to determine broad-band integrated spectra of a large
number of stars. For 
instance, models need to address questions such as what 
sets the spectral range of the coherent emission?  Why does the
spectrum turn over at low frequency?  Is it a spatial
scale (which could limit maser growth, for instance)?  Is it 
a lack of plasma at the right density, or a lack of turbulent drivers
in that plasma (either of which could limit plasma turbulent
emission)?

\subsection{Mean Profiles}

The mean profile reveals the structure of the emission region.  It
tells us about the geometry and density structure in the region, as
well as the local magnetic field (which is revealed through its
polarization signature).  When a broad frequency range can be studied,
the mean profiles in many stars turn out to show strong 
frequency evolution, as illustrated in Figure 2.  As we noted above,
the data often disagree with what the standard model predicts.

\begin{figure}[ht]
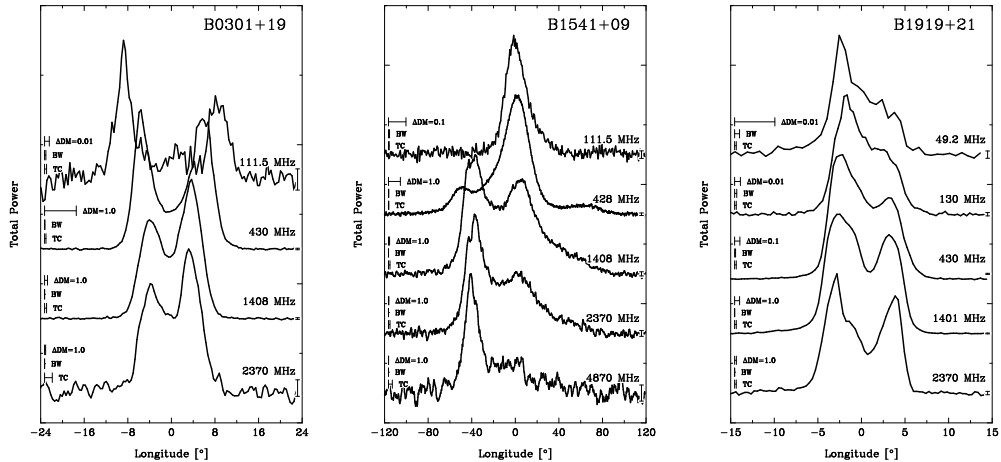


\begin{center}
\scalebox{0.26}{\includegraphics{Eilek2fig2a.ps}}
\hspace{2ex}
\scalebox{0.26}{\includegraphics{Eilek2fig2b.ps}}
\hspace{2ex}
\scalebox{0.26}{\includegraphics{Eilek2fig2c.ps}}
\end{center}
\vspace{-2ex}
\caption{Time-aligned mean profiles of three pulsars, illustrating 
  the complexity of the changes with frequency, and the importance of
  low-frequency observations in understanding each   star.  B0301+19
  agrees   with much of the standard model:  it    shows an emission
  cone which broadens at lower frequencies, and well 
  behaved polarization.  B1541+09 changes from an emission core at low
  frequencies to a one-sided emission cone at high frequencies;  its
  polarization strongly suggests a non-dipolar magnetic field in the
  emission region.  B1919+21 deviates from the standard model both in
  its mean profile and its polarization;  it was the first pulsar
  discovered and is still one of the least understood.  Data taken at Arecibo,
  1988-1990, from Hankins \& Rankin (2005), who also show polarization
  for each star.  }

\end{figure}

In order to understand the pulsar emission geometry in the important,
``non-standard'' stars, we need high quality mean profiles with
polarization information.  We also need 
the broadest possible frequency range, 
which can be reached by combining LWA data with higher-frequency data
from other telescopes.   Once again, such data will be important to
lead and test future models.  For instance, does the emission geometry
of the star change at frequencies below the spectral break?  Why do
some pulsars become strongly linearly polarized at low frequencies,
and does the emission geometry change with the polarization?  How
often does the linear polarization direction change with frequency
(which should not happen at all in the standard model)?

Because the pulsar signal can be integrated synchronously to form the
final profile, we anticipate that many stars will be bright enough to
be studied with milliperiod  resolution at good S/N.  We do note that
some stars will be too strongly broadened by interstellar
scintillation at these low frequencies;  but dispersion
measure is a rough measure of turbulent broadening ({\it e.g.},
L\"ohmer {\it et al.\,}2004), and quite a few stars have
sufficiently low  dispersion to be profitably studied even at tens of MHz.

\subsection{Single Pulse Studies}

This is the most challenging experiment but it will be worth the
effort. Single pulse studies at high time resolution tell us about the
time dependence of the plasma --- as reflected in such phenomena as
microstructure, rapid polarization fluctuations and drifting
subpulses.  As we have emphasized before, studies over a broad
frequency range are needed to understand a star and test the models.
Low frequencies are particularly interesting here,  because
microstructure and subpulses tend to be stronger at low frequencies,
where we would expect the density imbalance and plasma dynamics to be
the most noticable.

\begin{figure}[!ht]
\begin{center}
\hbox{
\scalebox{0.675}{\includegraphics{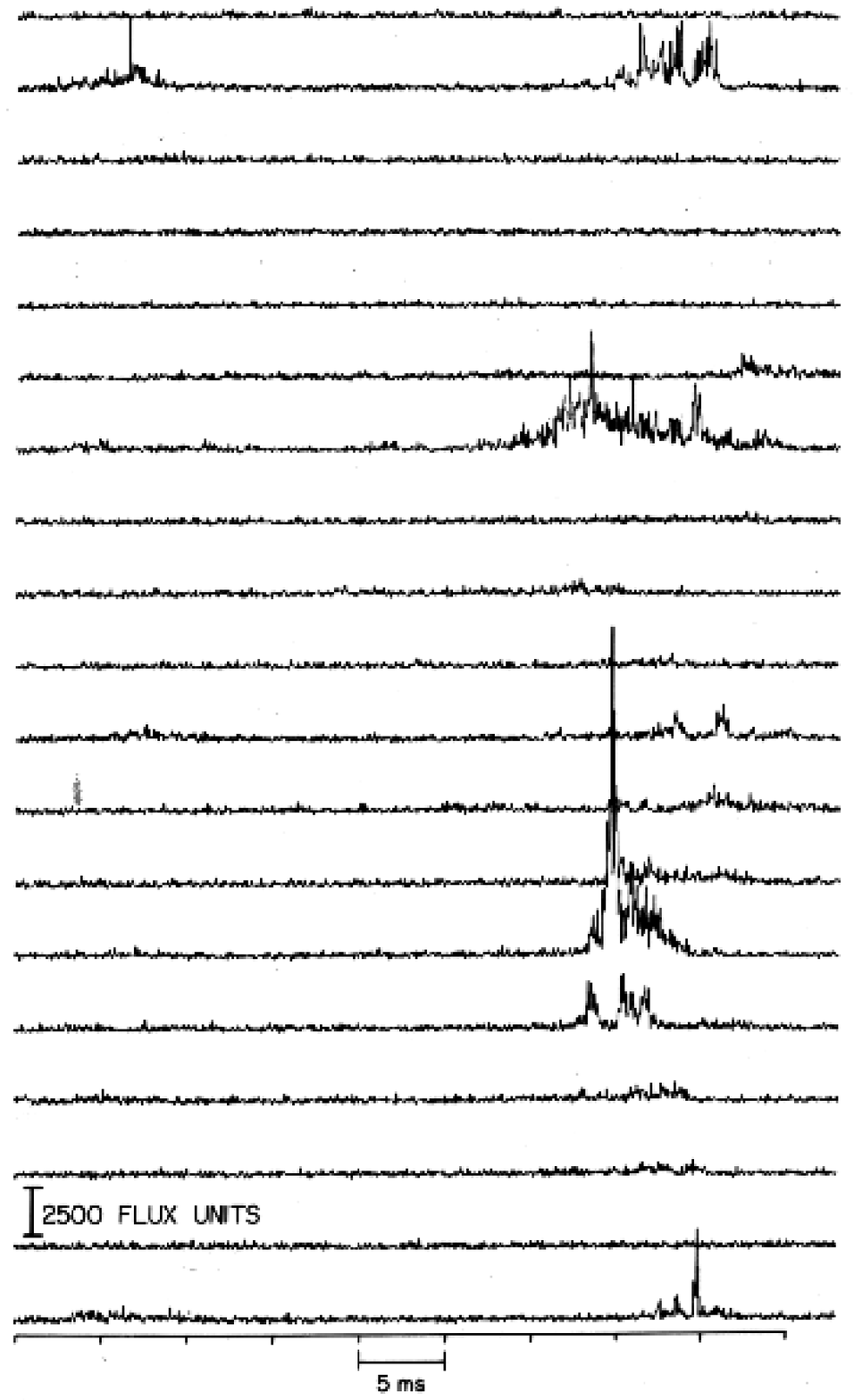}}
\hspace{4ex}
\rotatebox{90}{
\scalebox{0.695}{\includegraphics{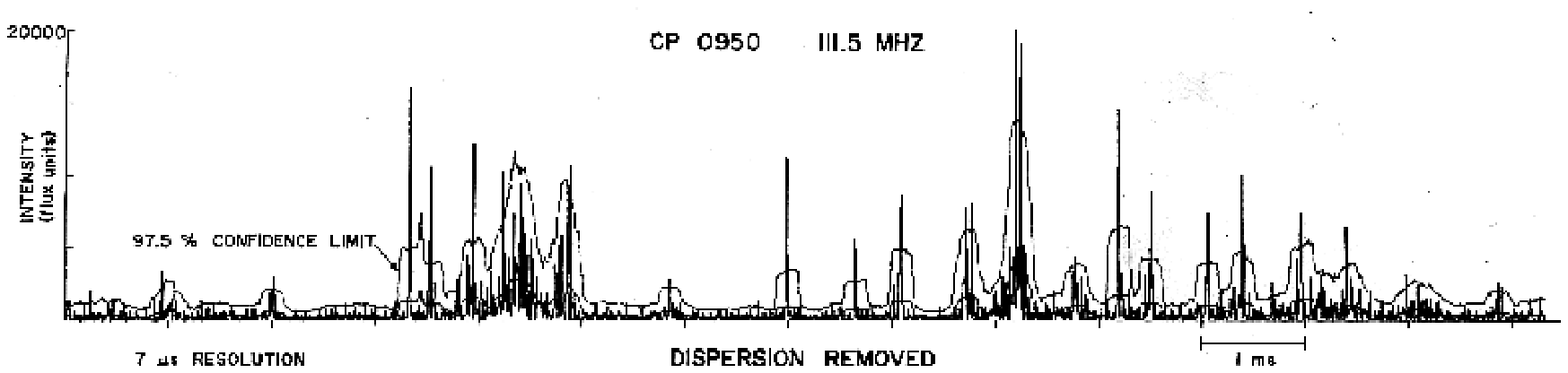}} }
}
\hspace{-6ex}
\end{center}
\vspace{-4ex}
\caption{Single pulses can vary tremendously in power, and when
  observed at high time resolution are found to contain
``giant microbursts'' with peak flux much larger than the pulsar's
  time-averaged flux.  Left, a sequence of single pulses from
  B1133+16, plotted at 56 microseconds time resolution.  Right, one
  pulse from B0950+08, at 7 microsecond time
  resolution with 97.5\% confidence limits shown.  Both observations
  at 111.5 MHz, from Hankins (1971). 
  }
\end{figure}

Observations such as these have
potentially high payoff, because they come closest to predictions that
can be made by theoretical models of the emission region.  They will
be difficult at low frequencies, however, because  sub-millisecond time
resolution is necessary which makes sensitivity and scattering
broadening  important issues. We believe a good number of bright puslars 
will be observable in this mode.  By current estimates, the LWA
will have thermal noise $\la 1 $mJy in one hour's integration, around
the middle of its frequency range.  This scales to $\la 3$Jy for
millisecond time resolution;  so single pulse work at high S/N can
only be done on pulsars as bright as, say, 300 Jy.  While this is
large compared to the  time-integrated flux for most pulsars, our
experience is that many pulsars show occasional very bright
microbursts.  Figure 3 shows two examples, each of which become
brighter than 10 kJy for brief moments, and thus easily detectable in
a triggered observation.

\section{Closing comments}

There is much still to be learned about the physics of pulsar radio
emission.  New observations have made it clear that the simple model,
which has been around nearly since pulsars were discovered, does not
supply all the answers.  We hope that the data to come from new
instruments can inspire a new understanding of the physics, not to
mention closer contact between observers and theorists.  

Low-frequency information is particuarly important here. Although the
LWA has been envisioned primarily as an imaging 
instrument, it can also make important contributions to pulsar
science, by providing critical low-frequency information which is not
easily available anywhere else. 

\acknowledgements 
We thank Don Backer and Frazer Owen for technical discussions, and the
members of the Socorro pulsar group for ongoing discussions on pulsar
physics. This work was partly supported by NSF grant AST-0139641.

\end{document}